\documentclass[12pt]{article}
\usepackage{graphicx}% Include figure files
\usepackage{scicite}
\usepackage{times}
\usepackage{amssymb}
\newenvironment{sciabstract}{%
\begin{quote} \bf}
{\end{quote}}

\newcounter{lastnote}
\newenvironment{scilastnote}{%
\setcounter{lastnote}{\value{enumiv}}%
\addtocounter{lastnote}{+1}%
\begin{list}%
{\arabic{lastnote}.}
{\setlength{\leftmargin}{0.85in}}
{\setlength{\labelsep}{.5em}}}
{\end{list}}

\title{1D-1D Coulomb Drag Signature of a Luttinger Liquid}

\author{D. Laroche$^{1,2}$, G. Gervais$^{1\ast}$, M. P. Lilly$^{2}$, J. L. Reno$^{2}$\\
\\
\normalsize{$^{1}$Department of Physics, McGill University, Montreal, H3A 2T8 CANADA}\\
\normalsize{$^{2}$Center for Integrated Nanotechnologies, Sandia National Laboratories, Albuquerque, NM 87185 USA}\\
\\
\normalsize{$^\ast$Corresponding author: gervais@physics.mcgill.ca}}

\date{}

\begin{document}


\baselineskip24pt


\maketitle

 \begin{sciabstract}

One-dimensional (1D) interacting electronic systems exhibit distinct properties when compared to their counterparts in higher dimensions.  We report Coulomb drag measurements between vertically integrated quantum wires separated by a barrier only 15 nanometers wide.  The temperature dependence of the drag resistance is measured in the true 1D regime where both wires have less than one 1D subband occupied.  As a function of temperature, an upturn in the drag resistance is observed below a temperature $ T^* \sim 1.6$ kelvin.  This crossover in Coulomb drag behavior is consistent with Tomonaga-Luttinger liquid models for the 1D-1D drag between quantum wires.

\end{sciabstract}
%   In addition to the interest of fundamentally understanding one-dimensional electron-hole asymmetry \cite{Kamenev}, energy fluctuations \cite{Sanchez} and the Tomonaga-Luttinger liquid (TLL) theory \cite{Tomonaga, Luttinger}, its exotic predictions such as spin-charge separation \cite{photo1,spin-charge,Pepper} and charge fractionalization \cite{fractionalization}, understanding the physics of one-dimensional system could be key in the development of self-powered nanodevices and might alleviate the problem of heat removal in densely packed electronic circuits \cite{Sothmann}.

 The exact role played by electron-electron interactions is greatly influenced by the dimensionality of the system.
 % For example,
 %in the one-dimensional limit (1D) where the exchange of particles is forbidden, Bose and Fermi statistics become obselete.
 In three dimensions, many interacting fermionic systems are well described by Landau's Fermi liquid (FL) theory, in which `quasiparticles'  are the elementary mean-field excitations of the system. In the strict one-dimensional (1D) limit, however, any perturbation in the motion of a single particle will affect all others, so mean-field (FL) theory is not applicable.  A Hamiltonian with exact analytic solutions can be constructed for a 1D quantum fluid \cite{Tomonaga, Luttinger,Haldane:1981gd}; however, the experimental realization of this so-called Tomonaga Luttinger liquid (TLL) in solid-state materials has proved to be difficult. TLL behavior may have been observed in correlated materials such as quasi-1D organic conductors and carbon nanotubes \cite{carbon1,carbon2}. In clean semiconductors, where the disorder can be kept at an extremely low level, previous 1D-1D  and 1D-2D tunneling experiments have shown strong evidence for spin-charge separation \cite{spin-charge,Pepper} and charge partition \cite{fractionalization},  as predicted by TLL models of semiconductor quantum wires. Here, we report on a 1D-1D Coulomb drag resistance that, below a crossover temperature $T^*$, increases with decreasing temperature. This upturn in the drag signal is consistent with predictions from Tomonaga-Luttinger theory of 1D-1D drag \cite{Stern1,Pustilnik}.

The Coulomb screening by free carriers in semiconductors is weak, and so the interaction between charges is long ranged. When two parallel conducting wires are separated by a small insulating barrier, a current in one wire generates a net charge displacement in the other so that a voltage develops by virtue of electron-electron interactions alone. The resistivity of this so-called Coulomb drag, defined
as $R_{D} \equiv - V_{drag} / I_{drive}$, where $I_{drive}$ is the drive current and $V_{drag}$ is the drag voltage, follows a $T^{2}$ dependence at low temperatures in 2D systems \cite{Gramilda}, in accordance with FL theory.

In 1D systems, however, FL theory is expected to break down because of the increased correlations and the Peierls instability. Several theoretical studies have explored the effects of electron-electron interactions on the drag resistance of coupled 1D-1D systems in terms of momentum transfer and TLL theory, with both backscattering\cite{Stern1, Lehur, Stern2, Averin1, Averin2, Flensberg, Dmitriev} and forward scattering \cite{Pustilnik} contributing to the drag signal.  These theories predict a positive drag signal between two wires with negative charge carriers (i.e. $R_{D} \geq 0 $), and models that include contributions from forward scattering predict an upturn in the drag at a temperature  $T^{*}$. Alternatively, 1D-1D Coulomb drag can also be described in terms of rectification of energy fluctuations \cite{Kamenev,Sanchez}, where both positive and negative drag signals can occur with a monotonic temperature dependence. In all cases, the strength of the drag signal will be influenced by a variety of parameters such as the 1D electron density, the interwire separation,  the number of 1D subbands, and the mismatch in the 1D electron densities.

Most experiments measuring 1D-1D Coulomb drag have been realized in a lateral geometry, where the interwire separation is large ($d$ $\gtrsim$ 200 nm) and where the barrier between the wires is soft (electrostatic) \cite{Debray,Yamamoto}. To circumvent the shadowing effect of the gates and so to decrease the interwire distance, a vertical design can be implemented (Fig. 1) in which the two wires are separated by a hard barrier defined by molecular beam epitaxy \cite{Laroche}. This approach results in two tunable 1D systems, each with their own source-drain contacts \cite{note}. We investigate the temperature dependence of the drag signal in such a system in a regime where less than one 1D subband is occupied in each wire.

%vertically coupled \cite{Laroche} quantum wires found the presence of a negative Coulomb drag signal, in apparent disagreement with the TLL momentum-tranfer for Coulomb drag and suggesting that perhaps more than a single mechanism is required to describe one-dimensional Coulomb drag over the large phase-space over which it is observed.  In an attempt to clarify the mechanisms at play in the true 1D-1D regime,
\begin{figure} [h]
\begin{center}
\includegraphics[width = 14cm]{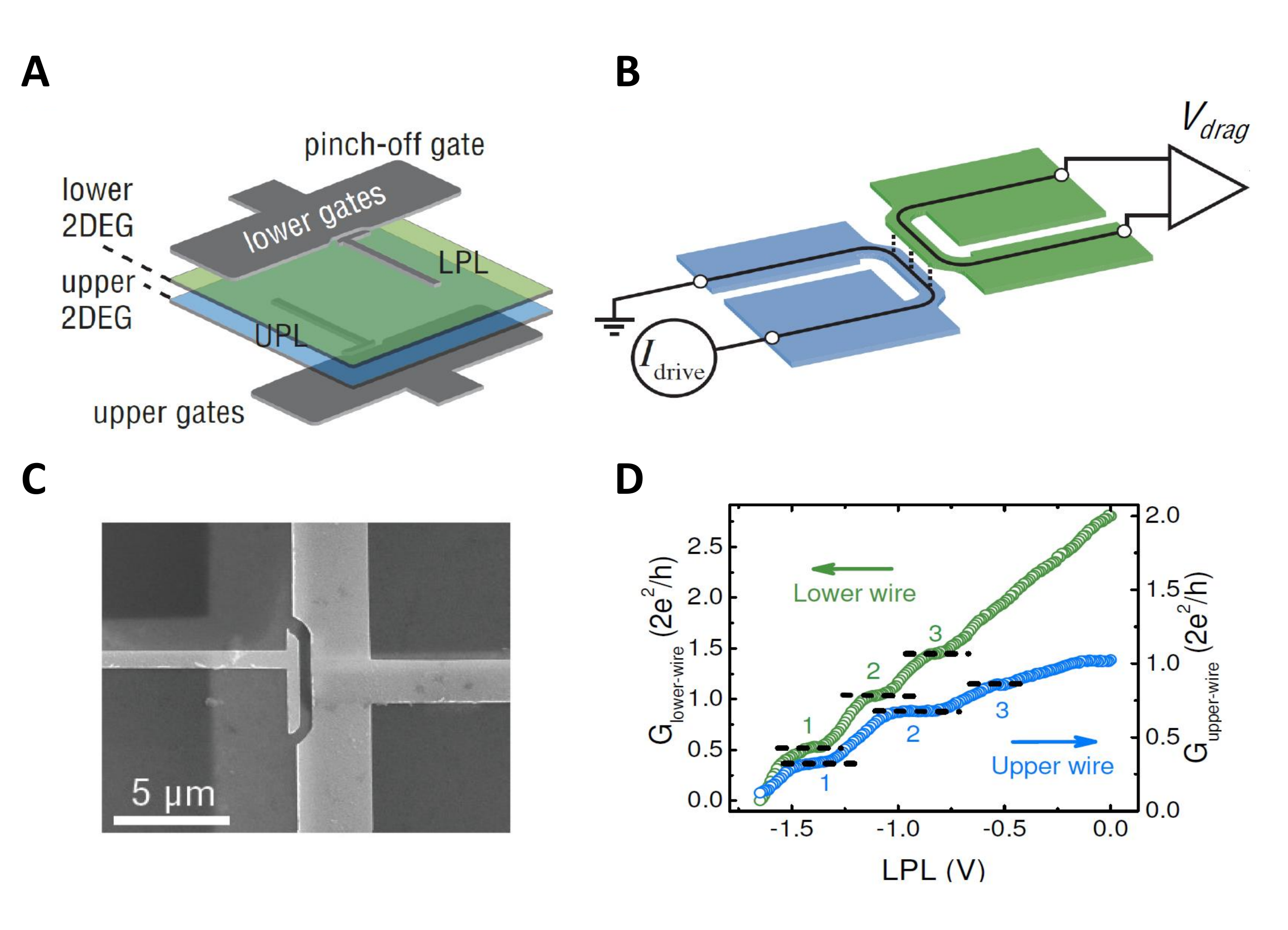}

\caption{\label{fig1} {\bf Design of the vertically integrated quantum wire device}. \small{\textbf{(A)} Schematic of the active part of the double quantum wires device.  The epoxy-bond-and-stop-etch (EBASE) process causes the lower gates and two-dimensional electron gas (2DEG) to be above the upper gates and 2DEG. \textbf{(B)}  In the interacting region of the device, two independent quantum wires are created, and superimposed vertically, and Coulomb drag measurements are performed.  \textbf{(C)} Scanning electron micrograph of the device.  The lower plunger (LPL) and pinch-off gate are visible on top of the device. The upper plunger (UPL) and pinch-off gate are also visible underneath the lower gates. After processing, the electron density in the upper (lower) layer is 1.1 (1.4) $\times 10^{11}$ cm$^{-2}$.  \textbf{(D)}  Typical conductance data of the lower quantum wire (green curve, left axis) and of the upper quantum wire (blue curve, right axis) from sample 2-C for fixed UPL = -0.23 V.  Because each gate is capacitively coupled to both wires, varying the voltage in a single gate affects the conductance of both wires.}}
\end{center}
\end{figure}

Typical examples of Coulomb drag measurements as a function of gate voltage are presented in Fig. 2, A and B, alongside conductance measurements in both wires for sample 2-L and 2-C, respectively.
Steps are observed in the conductance of the wires, but as the wires are non-ballistic, the conductance is not exactly quantized in units of $2e^{2}/h$, where $e$ is the elctronic charge and $h$ is Planck's contant. Such plateau-like features at reduced conductance $G<2e^{2}/h\times N$, with $N$ the number of quantum-mechanical channels, have been observed previously \cite{Yacobywire2}, and it was found that well-defined 1D subbands were still formed in the wires. Three main features are observed in the 1D-1D drag \cite{Laroche}: (i) peaks in the drag signal concomitant with the opening of 1D subbands, (ii) negative Coulomb drag at low density when the conductance in the drag wire is nearly depleted, and (iii) a negative Coulomb drag occurring between peaks in the drag signal (at a higher subband occupancy).  We reproduced these qualitative features of the drag signal in several devices over numerous cooldowns (see, e.g., Fig. 2A); the features are consistent with standard Coulomb drag tests such as frequency and drive-current independence \cite{Laroche}. The wires' subband occupancy (or 1D density) has a tremendous impact on the general behavior of the temperature dependence of the Coulomb drag signal (Fig. 2C). We now focus on the regime in which only one single 1D subband is occupied in each wire to probe electron-electron interaction in the true 1D regime.

%In the Laudauer formalism, which is valid in the presence of quantum wires connected to FL leads, the conductance for quantum transport assumes the value $G = \frac{2e^{2}}{h} \sum_{i=1}^{N} T_{i}$ where $N$ is the number of conduction channels and $T_{i}$ is the electron transmission probability for each channel.  In the ballistic regime where the electron motion is unhindered, the electron are fully transmitted across every channel and the conductance is simply $G = \frac{2e^{2}}{h}N$.  In the typical vertically-integrated quantum wires structures, additional defects are present due to the invasive fabrication techniques, causing the 1D electron to be in a non-ballistic regime where $G < \frac{2e^{2}}{h}N$, as depicted in figure 1\textbf{d}.  Plateau-like features at reduced conductance have been previously observed \cite{Yacobywire2} in quantum wires and have been found not to affect their 1D nature, making us confident that well-defined 1D subbands are indeed observed in our quantum wires.

\begin{figure}[H]
\begin{center}
\includegraphics[width = 9cm]{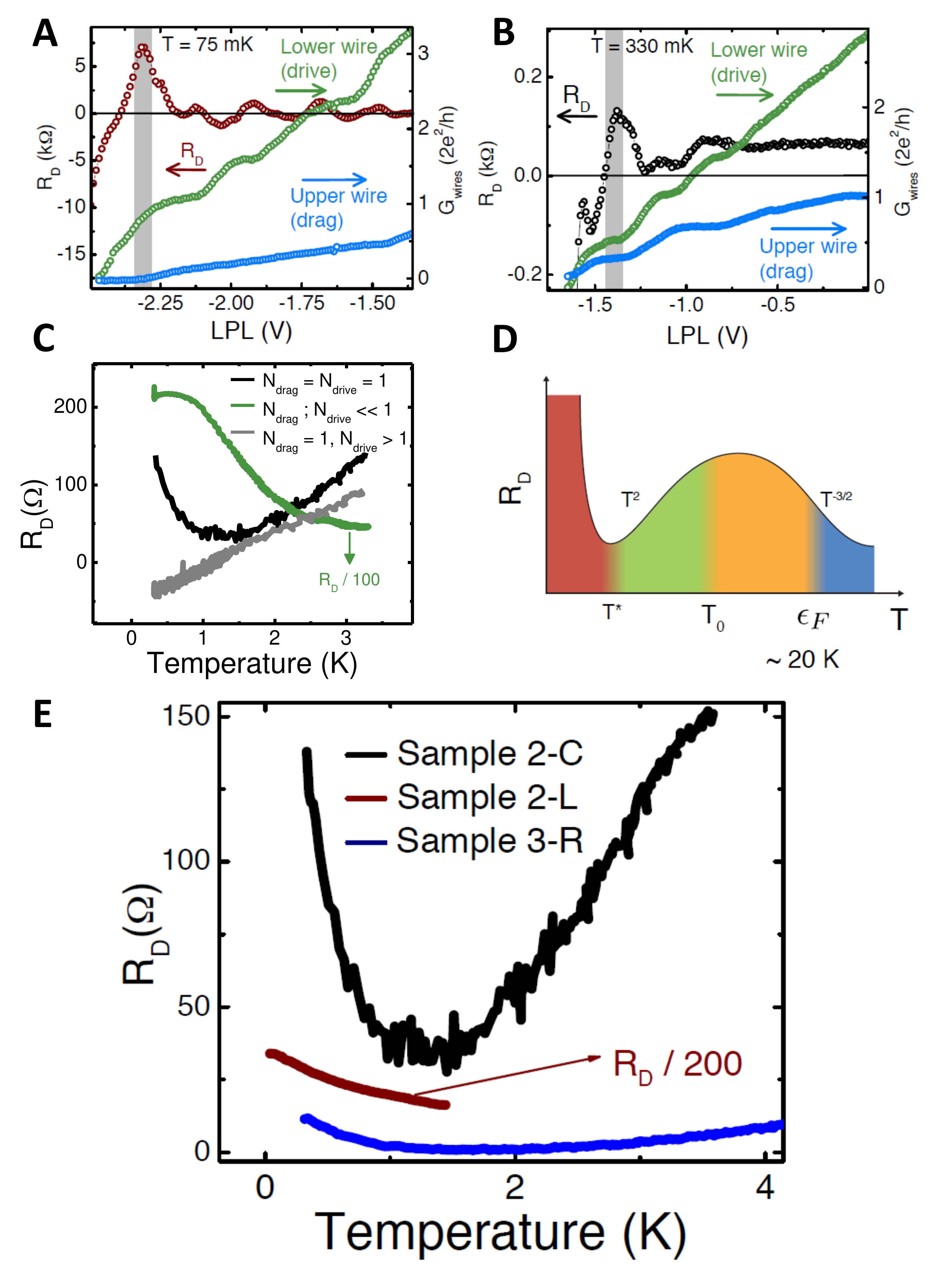}
\vspace*{-6mm}

\caption{\label{fig2} {\bf Coulomb drag measurements in vertically integrated quantum wires}.  \small{\textbf{(A)} Drag resistance at $T=75$ mK (red curve, left axis) versus gate voltage, along with the conductance in the upper and lower quantum wires (blue and green curves, respectively; right axis) for sample 2-L for fixed UPL = -0.15 V. \textbf{(B)} Drag resistance at $T=330$ mK (black curve, left axis) versus gate voltage, along with the conductance in the upper and lower quantum wires (blue and green curves, respectively; right axis) for sample 2-C for fixed UPL = -0.23 V.  \textbf{(C)} Temperature dependence of the Coulomb drag signal in sample 2-C for quantum wires with a single subband occupied (black curve), less than a full subband occupied (green curve), and slightly more than a single subband occupied (gray curve).  The drag signal changes drastically with the wire's subband occupancy, or 1D density.  \textbf{(D)} Expected behavior of the drag resistance versus temperature based on TLL theory, including corrections from forward scattering. [Cartoon reprinted with permission from \cite{Pustilnik}. Copyright (2003) by the American Physical Society.].  Here, $T_{0} = \hbar v_{F} d^{-1}$, $d$ is the interwire separation and $v_{F}$ is the Fermi velocity.  \textbf{(E)} Temperature dependence of the drag signal for samples 2-L, 2-C and 3-R.  For samples 2-L and 2-C, the temperature dependence was taken with no more than one 1D subband occupancy in each wires [highlighted by a gray stripe in (A) and (B), respectively], whereas the number of 1D subbands occupied in sample 3-R is bounded by $0 < N_{drive} \leq 2$ and $0 \leq N_{drag} \leq 3$. The magnitude of the drag resistance in sample 2-L is divided by 200 for visibility; the large difference in magnitude of the drag signal between devices is likely caused by slight differences in the density mismatch of the pair of wires from sample to sample, as $R_{D}$ is expected to decrease exponentially with increasing density mismatch.}}
\vspace{-5mm}
\end{center}
\end{figure}

Fig. 2E shows our main result, where the temperature dependence of 1D-1D Coulomb drag is shown when only a single 1D subband is present in both wires \cite{note2}.  Notably, a transition from a high-temperature regime where 1D Coulomb drag decreases with decreasing temperature, to a low-temperature regime where the Coulomb drag diverges as $T \rightarrow 0$, is observed.  This crossover occurs at a
temperature in the vicinity of $ T^{*} \simeq 1.6$ K in two samples, whereas the low-temperature behavior of the drag signal has been reproduced in three distinct samples. This upturn in the 1D-1D drag resistance is theoretically expected to occur for identical wires with relatively large interwire separation ($k_{F}d > 1$, where $k_{F}$ is the Fermi wave vector)\cite{Pustilnik, Lehur}. In a 1D system of length $L$, fluctuations
preclude the existence of any long-range order, and instead, the
electronic system should be described to lowest order by the TLL theory \cite{Haldane:1981gd} with
an effective Hamiltonian $H = H_{\rho} + H_{\sigma}$ separating charge ($\rho$) and spin components ($\sigma$) with
$$
H_{i} = (\hbar /2\pi)\int_0^L d z [(\partial_z \phi_{i})^2 u_{i}/K_{i} + K_{i}u_{i} (\partial_z \theta_{i})^2]
$$
%\begin{equation}
%H = \frac{\hbar v}{2\pi}\int_0^L d z \left[\frac{1}{K} \left(\partial_z \phi\right)^2 + K
%\left(\partial_z \theta\right)^2\right].
%\label{eq:LLHam}
%\end{equation}
where the phases $\phi(z)_{i}$ and $\theta(z)_{i}$  are defined in terms of
the second quantized electron field operator and $i = \rho, \sigma$ stands for the charge and the spin components, respectively. Its low-energy modes have dispersion $\varepsilon_{i}(k) = \hbar u_{i} k$ where $u_{i}$ is the velocity of the charge or spin component and is related to the underling many-body effective interaction parameters $K_{i}$. Assuming a system with spin-rotation symmetry  for which $K_{\sigma} = 1$, the Luttinger charge parameter $K_{\rho}$  tunes the system from attractive interactions ($K_{\rho} > 1$) to a system with repulsive interactions ($K_{\rho} < 1$).

Using a TLL model for identical wires \cite{Stern1} and accounting for the forward momentum-transfer corrections \cite{Pustilnik}, an upturn temperature $T^{*}$ was calculated to be $T^{*} \sim \epsilon_{F} e^{\frac{-k_{F}d}{1-K^{-}_{\rho}}}$, where $\epsilon_{F}$ is the Fermi energy. The many-body Luttinger liquid parameter $K^{-}_{\rho}$ here is the relative interaction parameter for antisymmetric charge displacement. It is defined as the difference between small intra- and interwire interaction parameters, yielding $K_{\rho}^-=1$ for Fermi-like systems. Within this model, backscattering should be the main source of the Coulomb drag signal for $ T < T^{*} $, yielding to the formation of interlocked charge density waves in the wires. At temperature much larger than $T^{*}$, forward scattering should dominate leading to a transition between an exponential dependence to a power-law dependence on temperature provided that $1 < k_{F}d  \sim 2.2$ and $L > L^{*} \sim \hbar v_{F} / T^{*} = 0.5$ $\mu$m, as is the case in our samples.  Here, $L^{*}$ is a critical length such that, for $L>L^{*}$, an exponential increase in drag resistance is expected as $T\rightarrow 0$.  Using our best estimate for the electronic density in the wires from $n_{1D}=\sqrt{n_{2D}}$, and an interwire distance $d\simeq40$ nm (corresponding to the barrier width plus half of both well widths, and the wire alignment uncertainty), we estimate $K_{\rho}^-\simeq 0.16 \pm 0.02$ ($K_{\rho}^-\simeq 0.08 \pm 0.02$) from the observed upturn temperatures in the $\simeq 1.3$ to $1.5$ K ($1.7$ to $1.9$ K) range for the wires in sample 2-C (3-R).  Although it is possible for backscattering alone to create such an upturn in the temperature dependence of Coulomb drag between identical wires when $K_{\rho}^- > 0.5$\cite{Stern1}, our wires appear to not be in this regime and also are not exactly matched.  For density-mismatched wires, backscattering alone can theoretically induce an upturn in the temperature dependence, regardless of the value of $K_{\rho}^-$, given a suitable density imbalance between the wires \cite{Stern2}.  Such an upturn has also been predicted to occur in the spin-incoherent regime of the Luttinger liquid where the spin exchange energy $J = \epsilon_{F}e^{-2.9/\sqrt{na_{b}}}$ is suppressed \cite{Lehur}, where  $a_{B} = \epsilon \hbar^{2}/m^{*}e^{2}$ is the Bohr radius, $\epsilon$ is the dielectric constant and $m^{*}$ is the electron effective mass.  This regime is expected to occur for $na_{B} \ll 1$ and $T^{*} < J < T < \epsilon_{F}$.  In our wires, we estimate $J \sim 150$ mK $ < T^{*}$ and $na_{B} \sim 0.4$, and therefore it is unikely that our data falls into this regime.

 %Le Hur  argued \cite{Lehur} that it should occur in  the vicinity of $T = J$, the spin-exchange energy, which is estimated to be $\sim 0.02$ mK in our sample following the model by Matveev \cite{Matveev}.  These predicted upturn temperatures are at least one order of magnitude lower than our experimentally observed upturn $T^{upturn} = 1.5$ K in the drag signal.

In Fig. 3, the drag signal is shown versus temperatures for samples 2-C and 3-R on a linear, a log-log, and an Arrhenius plot.  At temperatures $T>T^{*}$ , dependence of the drag signal appears consistent with a power law given by $R_{D}\sim T^{\gamma}$, with $\gamma$
given by $\gamma = 1.9 \pm 0.1$ and $\gamma = 3.0 \pm 0.1$ for sample 2-C and 3-R, respectively. This power law is expected from LL models, including corrections from forward scattering with a theoretical exponent $\gamma=2$. We caution, however, that this model considers Coulomb drag between wires with identical 1D densities. This is unlikely to be the case in our system because the parent electronic densities in the two 2D layers differ by
$\sim 20 \%$, and even a slight density imbalance of less than 2.5\%  between the wires
might have strong effect on the temperature dependence of the drag signal \cite{Pustilnik}. Defining a measure of density imbalance $T_{1} = k_{F}\delta(v)$, where $\delta(v)$ is the difference between the Fermi velocity in  both wires, the drag resistance is expected to be suppressed as $R_{D} \propto \frac{T}{T_{1}} e^{-T_{1}/T}$, provided $T_{1} \gtrsim T \gtrsim 450$ mK.  In our samples, the 1D density imbalance can be as large as 12 \%, giving  $T_{1} \sim 5.4$ K. Therefore, rather than a simple power-law dependence, the drag signal should behave as a convolution between this exponential decay and a power law, offering a pathway of explanation for the discrepancy between the exponent $\gamma$ in the two samples.  Extracting the experimental value of $T_{1}$ from a linear fit of the Arrhenius plots in the high-temperature regime, we obtain $T_{1} = 4.8 \pm 0.4$ K ($T_{1} = 10.7 \pm 0.4$ K) for sample 2-C (3-R).  These extracted $T_{1}$ values are comparable to the calculated $T_{1}=5.4$ K value from the estimated density imbalance in our wires.  Despite this apparent agreement with a TLL model for Coulomb drag, including forward-scattering corrections, we stress that we cannot entirely discriminate between scenarios involving backscattering alone because it also predicts an exponential decrease of $R_{D}$ with decreasing temperature for $T > T^{*}$ \cite{Stern2}.  Finite-length effects could also modify the temperature dependence of the drag resistance.  These effects are expected to be notable at low temperature and large drive bias voltage, i.e, for $u / \theta  = \frac{V_{drive}}{V_{L}} / \frac{T}{T_{L}} = eV_{drive} / T \gg 1$ where $V_{L}$ and $T_{L}$ are determined from the plasmon frequency of the system \cite{Peguiron}.  For  our wires, we estimate $u / \theta \sim 1.2 / T$, and thus finite-length effects are expected to become negligible for $T \gtrsim 0.6$ K, and so unlikely to modify the drag signal in the high temperature regime, as well as the observed upturn.

% As such, looking at the temperature dependence of sample 2-C and 3-R (Fig. 3), one would expect to observe the Coulomb drag temperature dependence to be linear at high temperature in a log-log plot (Fig. 3 \textbf{C)} and \textbf{D)} and linear at low temperature in an Arrhenius plot (Fig. 3 \textbf{E)} and \textbf{F)}).

%or an exponential divergence  \cite{Stern1, Pustilnik, Dmitriev}, depending on the temperature. An example of the theoretical temperature dependence of 1D Coulomb drag between identical wires calculated by Pustilnik \emph{et al.} \cite{Pustilnik} is shown in figure 2 \textbf{D}.

\begin{figure}[H]
\begin{center}
\includegraphics[width = 10cm]{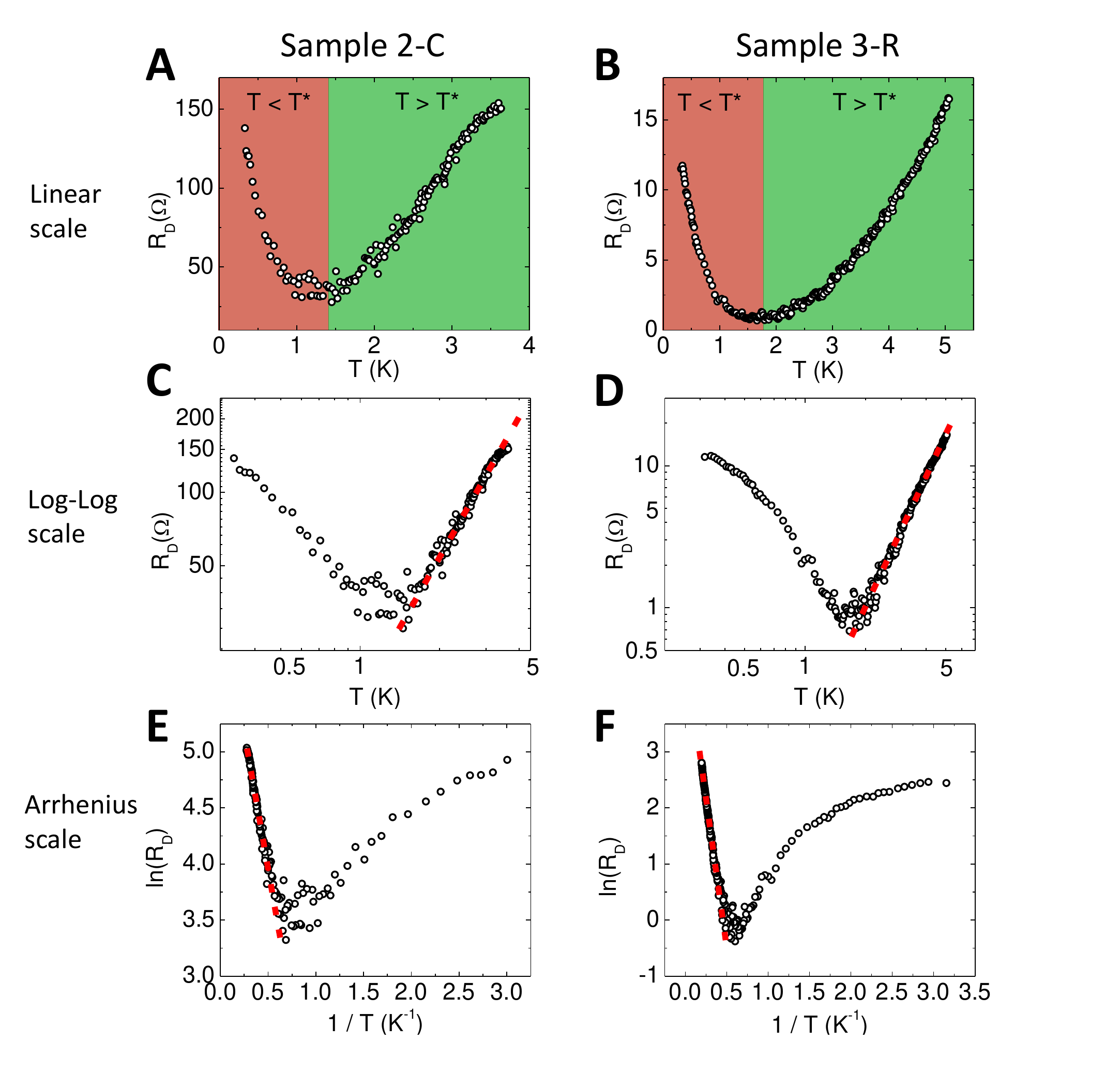}
\caption{\label{fig3}  {\bf Upturn in the drag resistance and TLL analysis}. \small{Drag resistance as a function of temperature for sample 2-C in a linear scale \textbf{(A)}, log-log scale \textbf{(C)}, and Arrhenius scale \textbf{(E)} and for sample 3-R in a linear scale \textbf{(B)}, log-log scale \textbf{(D)}, and Arrhenius scale \textbf{(F)}.  For both samples, the log-log and the Arrhenius scales appear to be linear at high temperature (red dotted line) and deviate from linearity at low temperature.  The analytic form is $R_{D} \propto a  T^{\gamma}$ for the log-log plot and  $R_{D} \propto b e^{-T_{1}/T}$ for the Arrhenius plot.  For sample 2-C, the fits yield, respectively,  $a = 14.5 \pm 1.3\Omega/K^\gamma$, $\gamma = 1.9 \pm 0.1$, and $b = 600 \pm 60\Omega$, $T_{1} = 4.8 \pm 0.4K$.  For sample 3-R, the fits yield $a = 0.12 \pm 0.01\Omega/K^\gamma$, $\gamma = 3.0 \pm 0.1$ and $b = 120 \pm 10\Omega$, $T_{1} = 10.7 \pm 0.4K$.  }}
\vspace*{-7mm}
\end{center}
\end{figure}

At temperatures below $T^{*}$, the drag signal is expected theoretically to diverge with decreasing T in the $T\rightarrow 0$ limit, with the exact form of the signal depending on the mismatch conditions between the wires \cite{Stern2}. This increase in drag resistance is a consequence of forward scattering dying out at the lowest temperatures and of algebraic decaying correlations of a Luttinger liquid. Although we unambiguously observe a drag signal increasing with decreasing temperature down to $T\simeq 75$ mK (Fig. 2E, sample 2-L), the present data do not allow us to extract the exact functional dependence upon temperature of the drag signal below $T^{*}$.  We also note that mesoscopic fluctuations and finite-size effects \cite{Peguiron,Mortensen, Narozhny} could contribute to a non monotonic temperature dependence of the drag resistance in the low-temperature regime. Future work is required to further explore the physics of 1D-1D drag in the  $T\rightarrow 0$ limit.

As is well known, the conductance of a quantum wire in the ballistic regime only possesses a very weak temperature dependence. On the contrary, the 1D-1D drag signal depends heavily on temperature, as well as on subband occupancy. Our observation of an upturn in the 1D-1D Coulomb drag signal confirms, at least qualitatively, an important prediction of Luttinger liquid models of quantum wires and potentially support theories accounting for the non linearity in the electron dispersion. The understanding of physics of interacting 1D systems is still in its infancy and as such, little is known regarding interacting Luttinger liquids. In the future, it may be possible to study in similarly fabricated devices interacting Luttinger liquids formed of interacting electron and holes, with different effective masses. Such devices might also be used to determine the existence of a nuclear spin helix, a recently predicted novel quantum state of matter \cite{Loss}. \\

\begin{scilastnote}
\item \textbf{Acknowledgements} We acknowledge the technical assistance of D. Tibbetts and J. Hedberg, and we thank I. Affleck for illuminating discussions. This work has been supported by the Division of Materials Sciences and Engineering, Office of Basic Energy Sciences, U.S. Department of Energy (DOE).  This work was performed, in part, at the Center for Integrated Nanotechnologies, a U.S. DOE, Office of Basic Energy Sciences, user facility.  Sandia National Laboratories is a multi program laboratory managed and operated by Sandia Corporation, a wholly owned subsidiary of Lockheed Martin Corporation, for the U.S. DOE's National Nuclear Security Administration under contract DE-AC04-94AL85000.  We also acknowledge the financial support from the Natural Sciences and Engineering Research Council of Canada (NSERC), the Canadian Institute For Advanced Research (CIFAR) and the Fonds Qu\'eb\'ecois de la Recherche sur la Nature et les Technologies (FQRNT).  All data and fabrication recipes presented in this work are available upon request to G.G.
\end{scilastnote}

\noindent \textbf{Supplementary Materials}\\
www.sciencemag.org\\
Materials and Methods\\
Received 2 August 2013; accepted 14 January 2014\\
Published online 23 January 2014

\clearpage

\section*{Supplementary Materials}
\noindent {\bf Materials and methods} \\
The vertically-integrated double quantum wires used in this Letter are patterned on a n-doped GaAs/AlGaAs electron bilayer heterostructure (wafer EA0975).  The two 18 nm wide quantum wells are separated by a 15 nm wide Al$_{0.3}$Ga$_{0.7}$As barrier.  The electron density is 1.1 (1.4) $\times 10^{11}$ cm$^{-2}$ for the upper (lower) 2DEG, yielding a combined mobility of $4.0 \times 10^{5}$ cm$^{2}$ / V$\cdot$ s.  Two sets of two split gates are deposited using electron beam lithography on the upper and the lower side of the sample using an epoxy-bound-and-stop-etch (EBASE) process. These gates define two wires that are $\sim 4.2$ $\mu$m long and $ \sim 0.5$ $\mu$m wide.  More details on the fabrication process is presented elsewhere \cite{Laroche}.  A schematics of the design used for the vertically-integrated quantum wires is presented in Fig. 1\textbf{A}, along with a SEM picture of a typical sample in Fig. 1\textbf{C}.  From the typical vertical alignment uncertainty observed in our structures, the center-to-center interwire separation is bounded between 33 nm and 41 nm. The three samples used in this Letter (sample 2-L, 2-C and 3-R) were fabricated from the same heterostrucure following identical processing steps.  Sample 2-L and 3-R were both measured in a $^{3}$He refrigerator at temperature down to 330 mK and sample 2-C was measured in a dilution refrigerator with an electron temperature reaching down to $\sim$ 75 mK.  The measurements are performed using standard low frequency lock-in amplifier techniques.  The conductance is measured using a two-wire measurements with an excitation voltage of 50 $\mu V$ at a frequency of 9 Hz and the Coulomb drag measurements are performed sourcing a 4.5 nA current at a frequency of 9 Hz.   Self consistency of the Coulomb drag measurement has also been verified (see \cite{Laroche} for more details).

The standard device operation goes as follow.  First both the lower and the upper pinch-off gates (LPO and UPO) are adjusted such that they deplete the 2DEG closest to them without depleting significantly the 2DEG farthest from them. This allows for the suppression of 2D tunnelling between the layers and to create independent contacts to each layer.  These gates are then held at constant bias for the remainder of the Coulomb drag experiment. Following this, the bias applied to the plunger gates (LPL and UPL for the lower and upper plunger gates, respectively) is adjusted to create a single, independently contacted quantum wire in each layer.  Since both plunger gates are capacitively coupled, sweeping the bias of a single gate affects the conductance of both quantum wires. Therefore, in a typical gate-voltage analysis experiment, the drag resistance is determined as a function of a single gate voltage (LPL in this report) while the other three gates are held at a fixed (negative) bias. A complete mapping of the drag resistance as a function of 1D subband occupancy can be obtained by performing successive LPL bias sweeps for all accessible UPL bias values, and these have published in an earlier report\cite{Laroche}.


\begin{thebibliography}{References}
\bibitem{note2} Because all devices were fabricated from the same parent heterostructure according to the same fabrication process, we assume that the 1D density of a given wire is roughly similar when tuned at a single 1D subband occupancy.
\end{thebibliography}
\end{document}